\documentclass[twoside]{article}

\usepackage{sequence_format}
\usepackage{verbatim}
\usepackage{graphicx}
\usepackage{xcolor}

\usepackage[numbers]{natbib}
\usepackage{microtype}

\usepackage[hide=false,setmargin=true,marginparwidth=0.5in]{marginalia}

\usepackage{multicol}
\usepackage{}

\usepackage{graphicx} % more modern
\usepackage{subfigure}

\usepackage{url}
\usepackage{bm}
\usepackage{framed}

\usepackage{amsmath}
\usepackage{amsfonts}
\DeclareMathAlphabet{\mathpzc}{OT1}{pzc}{m}{it}

\usepackage{stackrel}

\usepackage[shortlabels]{enumitem}

\usepackage{xcolor}
\usepackage{tikz}

\usepackage{rotating}

\usepackage{amsthm}
\usepackage{mathtools}

\usepackage{algorithm}
\usepackage{algpseudocode}

\usepackage[overlay]{textpos}

\theoremstyle{theorem}

\theoremstyle{lemma}

\theoremstyle{definition}

\theoremstyle{conjecture}

\usetikzlibrary{calc,trees,positioning,arrows,chains,shapes.geometric,%
    decorations.pathreplacing,decorations.pathmorphing,shapes,%
    matrix,shapes.symbols}

\tikzset{
>=stealth',
  punktchain/.style={
    rectangle, 
    rounded corners, 
    draw=black, very thick,
    text width=10em, 
    minimum height=3em, 
    text centered, 
    on chain},
  line/.style={draw, thick, <-},
  element/.style={
    tape,
    top color=white,
    bottom color=blue!50!black!60!,
    minimum width=8em,
    draw=blue!40!black!90, very thick,
    text width=10em, 
    minimum height=3.5em, 
    text centered, 
    on chain},
  every join/.style={->, thick,shorten >=1pt},
  decoration={brace},
  tuborg/.style={decorate},
  tubnode/.style={midway, right=2pt},
}

\usepackage{hyperref}

\usepackage{dblfloatfix}
\usepackage{float}
\usepackage{inconsolata}

\usepackage{hyperref}
\hypersetup{
  colorlinks=true,
  citecolor=blue,
  urlcolor=magenta,
}
\begin{document}

\twocolumn[
  \ourtitle{Rapid Generation of Stochastic Signals with Specified Statistics}
  \ourauthor{Span Spanbauer \And Ian Hunter}
  \ouraddress{MIT \And MIT}

%\vspace{-.5cm}
]

%%%%% %%%%% %%%%% %%%%% %%%%% %%%%% %%%%% %%%%% %%%%% %%%%% %%%%% %%%%%

\begin{abstract}
We demonstrate a novel algorithm for generating stationary stochastic signals with a specified power spectral density (or equivalently, via the Wiener-Khinchin relation, a specified autocorrelation function) while satisfying constraints on the signal's probability density function. A tightly related problem has already been essentially solved by methods involving nonlinear filtering, however we use a fundamentally different approach involving optimization and stochastic interchange which immediately generalizes to generating signals with a broader range of statistics. This combination of optimization and stochastic interchange eliminates drawbacks associated with either method in isolation, improving the best-case scaling in runtime to generate a signal of length $n$ from $\mathcal{O}(n^2)$ for stochastic interchange on its own to $\mathcal{O}(n \: \text{log} \: n)$ without parallelization or $\mathcal{O}(n)$ with full parallelization. We demonstrate this speedup experimentally, and furthermore show that the signals we generate match the desired autocorrelation more accurately than those generated by stochastic interchange on its own. We observe that the signals we produce, unlike those generated by optimization on its own, are stationary.

\end{abstract}

%%%%% %%%%% %%%%% %%%%% %%%%% %%%%% %%%%% %%%%% %%%%% %%%%% %%%%% %%%%%
\section{Introduction}
%%%%% %%%%% %%%%% %%%%% %%%%% %%%%% %%%%% %%%%% %%%%% %%%%% %%%%% %%%%%

Generating stochastic signals with specified statistics is a foundational problem with a long history. The problem has been motivated independently by myriad scientific applications including the generation of experimental stimuli for use in system identification~\citep{Godfrey1993PerturbationSF,tangorra2004stochastic}, for radar system testing~\citep{minfen2003method}, and for the simulation of specific non-Gaussian processes which occur in, for example, oceanography~\citep{brockett1987nonlinear}. The specific problem of generating a stationary stochastic signal with a specified autocorrelation and probability density function (PDF) has been largely solved via methods that involve sampling from specially designed Markov processes~\citep{rao1992generation,cai1996generation} including methods that involve nonlinear filtering of Gaussian white noise~\citep{gujar1968,liu1982generation,kontorovich1995stochastic}.

However, these methods are challenging to generalize. There is a single existing method which solves the problem of generating a stationary stochastic signal with a specific autocorrelation and PDF in full generality without nonlinear filtering: the stochastic interchange method by Hunter and Kearney~\citep{hunter1983generation}. This approach can be easily generalized to produce signals with a variety of properties, for example 2D stochastic signals with specified 2D autocorrelation and PDF, or signals with a specified higher-order autocorrelation and PDF. The primary drawback of this approach is speed; generating a signal of size $n$ via stochastic interchange requires at least $\mathcal{O}(n^2)$ time, whereas the nonlinear filtering methods only require $\mathcal{O}(n)$ time.

We present a method of accelerating the stochastic interchange method from a best-case runtime of $\mathcal{O}(n^2)$ to a best-case runtime of $\mathcal{O}(n \: \text{log} \: n)$ without parallelization or $\mathcal{O}(n)$ with full parallelization, while improving its accuracy and improving its capability to easily generalize to other statistics.

\noindent {\bf Contributions.} This paper presents a novel algorithm for generating stationary random signals with a specified autocorrelation function and constraints on the PDF. Specifically, it presents the following contributions:
\vspace*{-5pt}
\begin{enumerate}
\item This paper introduces a novel algorithm combining stochastic interchange with optimization steps, eliminating drawbacks associated with either method in isolation.
\item This paper presents experimental results showing that optimization plus stochastic interchange is faster than stochastic interchange on its own; the best-case scaling in input length $n$ for stochastic interchange is $\mathcal{O}(n^2)$, whereas optimization plus stochastic interchange can achieve $\mathcal{O}(n \: \text{log} \: n)$ without parallelization or $\mathcal{O}(n)$ with full parallelization.
\item This paper presents experimental results showing that optimization plus stochastic interchange generates signals matching the desired autocorrelation more accurately than stochastic interchange on its own.
\item This paper shows that optimization on its own can generate highly nonstationary signals, a problem which is resolved via the introduction of stochastic interchange steps.
\end{enumerate}

\begin{algorithm}[t!]
    \caption{Stochastic Interchange Method~\cite{hunter1983generation}}
    \begin{algorithmic}[1]
    	\State \textbf{Input:} Autocorrelation $A$
    	\State \hspace{11.3mm} Metric for comparing autocorrelations $D$
    	\State \hspace{11.3mm} PDF $P$
    	\State \hspace{11.3mm} Sequence length $n$
    	\State \hspace{11.3mm} Steps to run $T$
    	\State Generate $x_{1..n}$ with PDF $P$ via numerical integration of $P$ and inverse transform sampling.
        \State Compute $A_x$, the autocorrelation of $x$.
        \For {$t$ in $\{1, \ldots, T\}$}
        \State Draw $i$ and $j$ uniformly at random from $1..n$.
        \State Let $x'$ be $x$ with the $i$ and $j$ entries swapped.
        \State Compute $A_{x'}$, the autocorrelation of $x'$, in linear time (See \citep{hunter1983generation}).
        \If{$D(A,A_{x'})<D(A,A_{x})$}
        \State Let $x = x'$
        \EndIf
        \EndFor\\
        \Return $x$
    \end{algorithmic}
    \label{alg:interchange}
\end{algorithm}

%%%%% %%%%% %%%%% %%%%% %%%%% %%%%% %%%%% %%%%% %%%%% %%%%% %%%%% %%%%%
%\section{Background}
%\label{sec:background}
%%%%% %%%%% %%%%% %%%%% %%%%% %%%%% %%%%% %%%%% %%%%% %%%%% %%%%% %%%%%

\section{Background}

Hunter and Kearney~\cite{hunter1983generation} previously addressed the problem of generating signals with a desired PDF and autocorrelation via the stochastic interchange of samples as detailed in Algorithm~\ref{alg:interchange}. They begin by generating a signal with the desired PDF via numerical integration of the PDF and inverse transform sampling. They then repeatedly stochastically swap samples of the signal, only keeping the swap if the autocorrelation of the new signal is closer to the desired autocorrelation. These swaps retain the signal's PDF exactly while causing the signal's autocorrelation to monotonically converge towards the desired autocorrelation.

Stochastic interchange on its own is capable of generating moderate length input signals with a desired autocorrelation function and PDF. However it has two primary drawbacks: 
\begin{enumerate}
\item Generating a signal of length $n$ via stochastic interchange takes at least $\mathcal{O}(n^2)$ time, making it impractical to generate very long signals.
\item The stochastic interchange method requires the user to choose a specific PDF which is matched exactly by the algorithm; in the context of system identification it is preferable to instead set constraints on the PDF. Choosing a specific PDF is typically over-constraining and prevents the autocorrelation of the signal from matching the desired autocorrelation closely. This also limits the generalizability of the algorithm to a broader set of statistics.
\end{enumerate}

Both of these drawbacks can be resolved by replacing stochastic interchange with an optimization procedure. However, we observe that plain optimization also has drawbacks:
\begin{enumerate}
\item Optimization on its own often gets stuck in local optima. This is particularly common when the constraints on the PDF are tight in the context of achieving a desired autocorrelation function, which for example occurs commonly when a high signal power is desired in a limited signal range.
\item Optimization on its own often produces an obviously nonstationary signal as shown in Section~\ref{sec:results}; this is problematic if the goal is to match a desired power spectral density, since the Wiener-Khinchin relation which links power spectral density and autocorrelation requires stationarity.
\end{enumerate}

\begin{algorithm}[t!]
    \caption{Optimization + Stochastic Interchange}
    \begin{algorithmic}[1]
    	\State \textbf{Input:} Autocorrelation $A$
    	\State \hspace{11.3mm} Metric for comparing autocorrelations $D$
    	\State \hspace{11.3mm} Loss $L(x)$ establishing a constraint on PDF
    	\State \hspace{11.3mm} Sequence length $n$
    	\State \hspace{11.3mm} Steps to run $T$
        \State Generate $x_{1..n}$ independent samples from a Gaussian distribution.
        \For {$t$ in $\{1, \ldots, T\}$}
        \State Compute $A_x$, the autocorrelation of $x$.
        \State Compute the loss $D(A,A_x)+L(x)$ and update $x$ via a gradient step.
        \State Draw $i$ and $j$ uniformly at random from $1..n$.
        \State Let $x'$ be $x$ with the $i$ and $j$ entries swapped.
        \State Compute $A_{x'}$, the autocorrelation of $x'$.
        \If{$D(A,A_{x'})<D(A,A_{x})$}
        \State Let $x = x'$
        \EndIf
        \EndFor\\
        \Return $x$
    \end{algorithmic}
    \label{alg:new}
\end{algorithm}

\begin{figure*}[!t]

\centering
\vspace{-16mm}
\begin{picture}(320,170)
\put(0,0){\includegraphics[width=110mm]{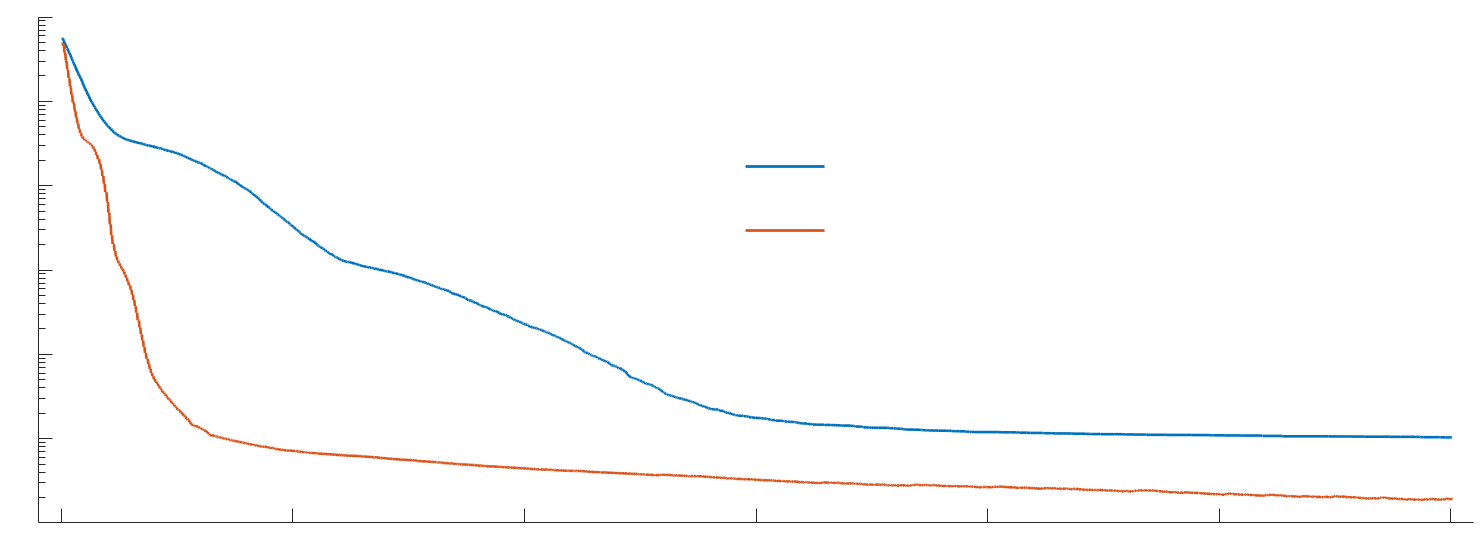}}
\put(20,118){\textbf{Comparison of convergence rate (10k sample signal)}}
\put(10.5,-5){0}
\put(57,-5){50}
\put(103,-5){100}
\put(152,-5){150}
\put(201,-5){200}
\put(250,-5){250}
\put(299,-5){300}
\put(-11,2){10\textsuperscript{-7}}
\put(-11,19){10\textsuperscript{-6}}
\put(-11,37){10\textsuperscript{-5}}
\put(-11,55){10\textsuperscript{-4}}
\put(-11,72){10\textsuperscript{-3}}
\put(-11,90){10\textsuperscript{-2}}
\put(-11,108){10\textsuperscript{-1}}
\put(143,-22){Time(s)}
\put(-26,45){\rotatebox{90}{$L_2$ loss}}
\put(180,78){Interchange alone}
\put(180,64){Optimization \& Interchange}
\end{picture}

\vspace{-2mm}
\begin{picture}(320,170)
\put(0,0){\includegraphics[width=110mm]{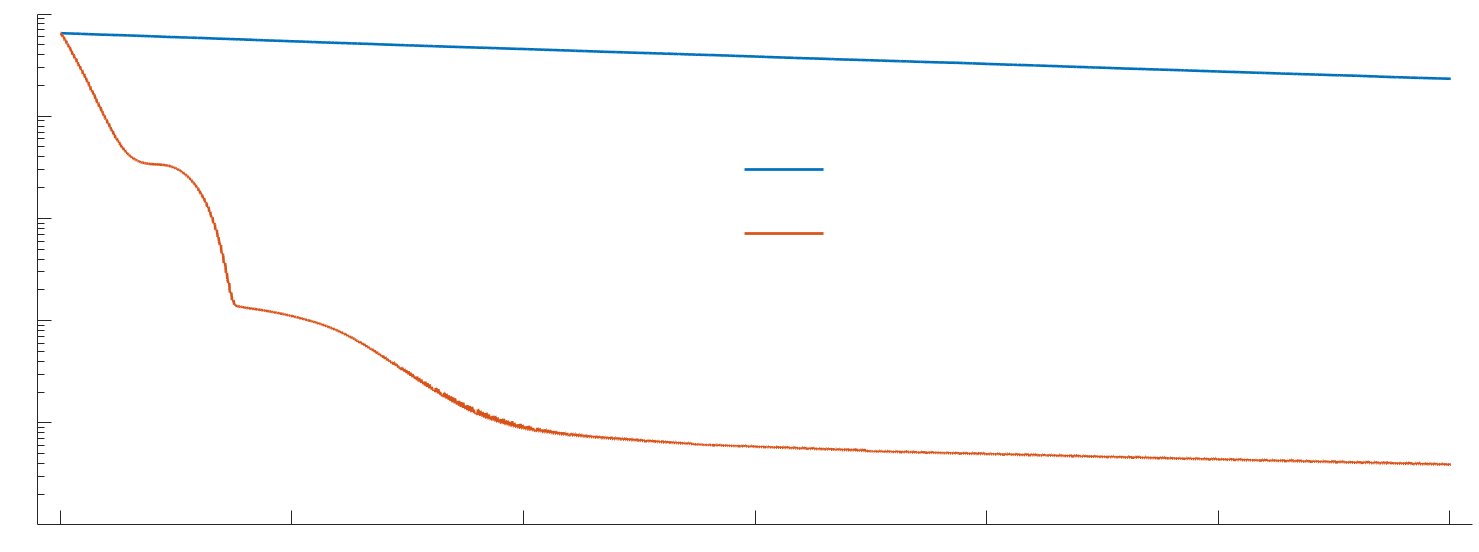}}
\put(20,118){\textbf{Comparison of convergence rate (100k sample signal)}}
\put(10.5,-5){0}
\put(57,-5){50}
\put(103,-5){100}
\put(152,-5){150}
\put(201,-5){200}
\put(250,-5){250}
\put(299,-5){300}
\put(-11,2){10\textsuperscript{-6}}
\put(-11,23){10\textsuperscript{-5}}
\put(-11,45){10\textsuperscript{-4}}
\put(-11,67){10\textsuperscript{-3}}
\put(-11,88){10\textsuperscript{-2}}
\put(-11,110){10\textsuperscript{-1}}
\put(143,-22){Time(s)}
\put(-26,45){\rotatebox{90}{$L_2$ loss}}
\put(180,78){Interchange alone}
\put(180,64){Optimization \& Interchange}
\end{picture}

\vspace{8.5mm}

\caption{Comparison of the convergence rate of the signal autocorrelation to the desired autocorrelation. To generate a signal of length $n$, the stochastic interchange method on its own takes at least $\mathcal{O}(n^2)$ time, whereas optimization plus stochastic interchange can take as little as $\mathcal{O}(n \: \text{log} \: n)$ time without parallelization or $\mathcal{O}(n)$ time with full parallelization. Here we see that optimization plus stochastic interchange converges much more quickly than stochastic interchange on its own, an advantage that grows with signal length. We also observe, in the 10k sample length case, that optimization plus stochastic interchange is able to reduce the variance of the difference between the signal autocorrelation and the desired autocorrelation by an order of magnitude beyond that of stochastic interchange on its own. This is because the optimization method has additional flexibility: the ability to adjust the PDF of the signal under constraint.}
\label{fig:convergence}
\end{figure*}
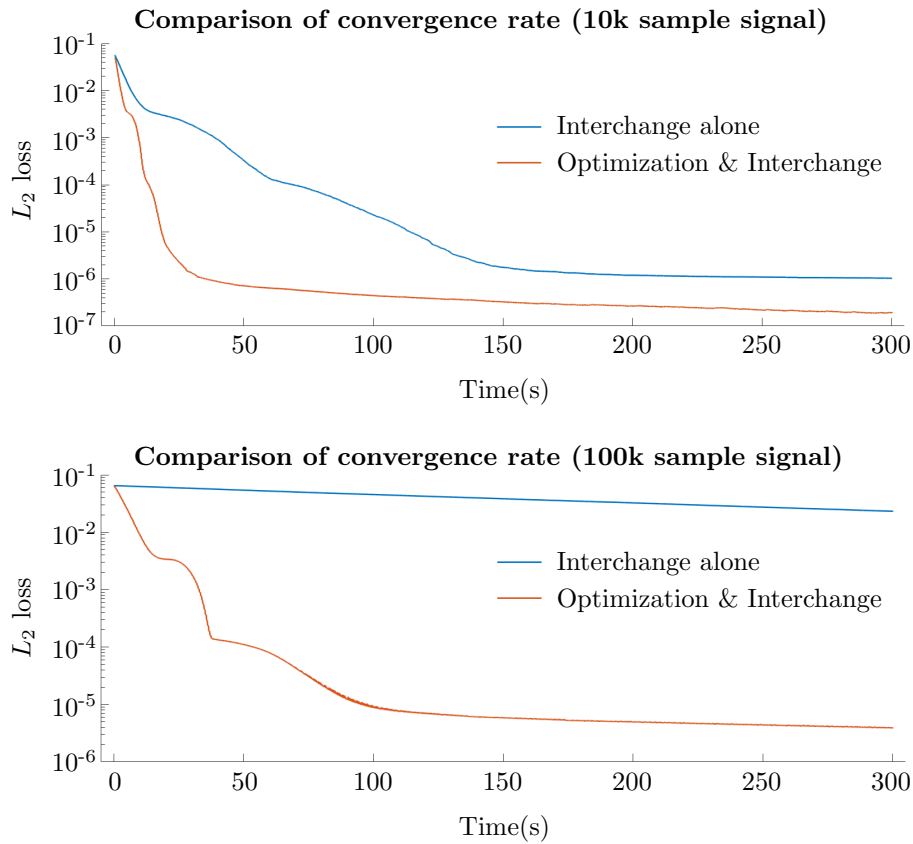

\section{Algorithm}
We resolve the aforementioned drawbacks by interspersing stochastic interchange steps with optimization steps as detailed in Algorithm~\ref{alg:new}. 

Instead of choosing a specific PDF we choose a loss function enforcing constraints on the PDF while simultaneously penalizing deviation from the desired autocorrelation. This is desirable in the context of system identification: if the goal of choosing a specific PDF is to constrain the range of inputs so that they can be generated by a certain actuator, then choosing appropriate constraints instead of a specific PDF allows additional flexibility in the signal, enabling the desired autocorrelation to be matched more closely; this effect can be seen in Fig.~\ref{fig:convergence}. If one does want to match both a specific PDF and autocorrelation, then optimizing a loss function mutually penalizing deviation from the desired PDF and the desired autocorrelation allows controlled tradeoff in the accuracy of the attained PDF and autocorrelation.

In our experiments we choose a loss function constraining the PDF to lie within the range [-0.5,0.5] with a linear penalty for signal values outside that range, and we choose to penalize deviations in the autocorrelation according to an $L_2$ norm.

We now turn to comparing the performance of optimization plus stochastic interchange with the performance of optimization on its own and with the performance of stochastic interchange on its own. We consider the problem of generating a signal of 10k samples and of 100k samples with a specified autocorrelation and a constrained PDF. All experiments were performed using an NVIDIA RTX 2080 Ti GPU. For gradient optimization steps we used an Adam optimizer~\citep{kingma2014adam} and the PyTorch~\citep{NEURIPS2019_9015} framework.

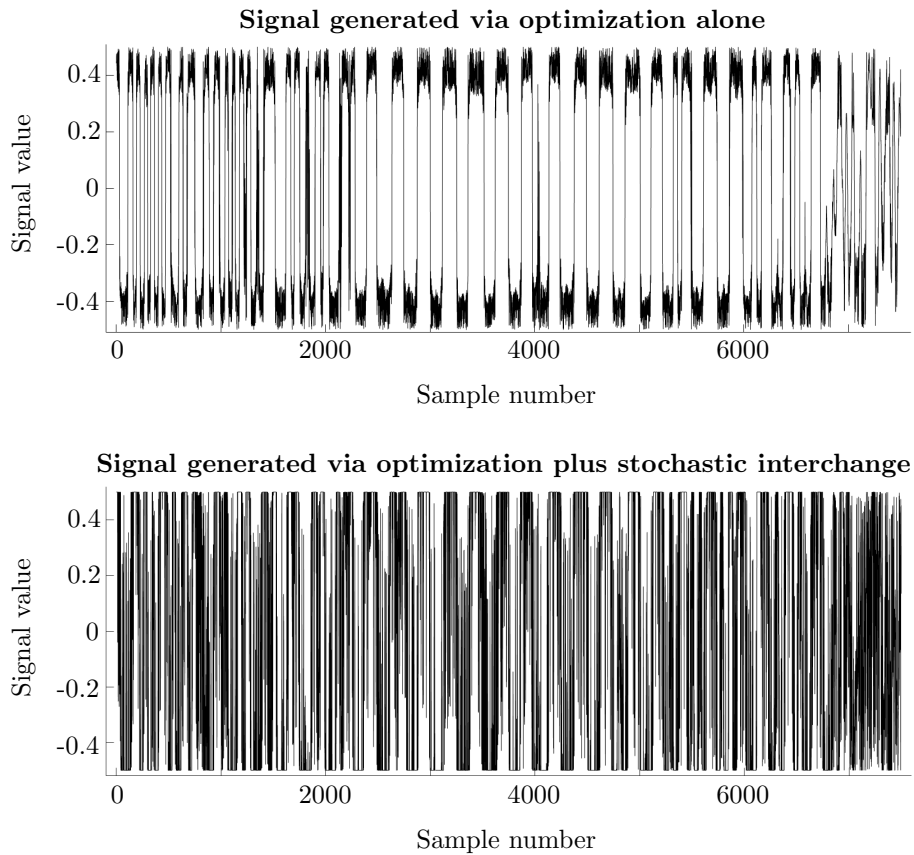
\begin{figure*}[!t]

\centering
\vspace{-15mm}
\begin{picture}(320,170)
\put(0,0){\includegraphics[width=110mm]{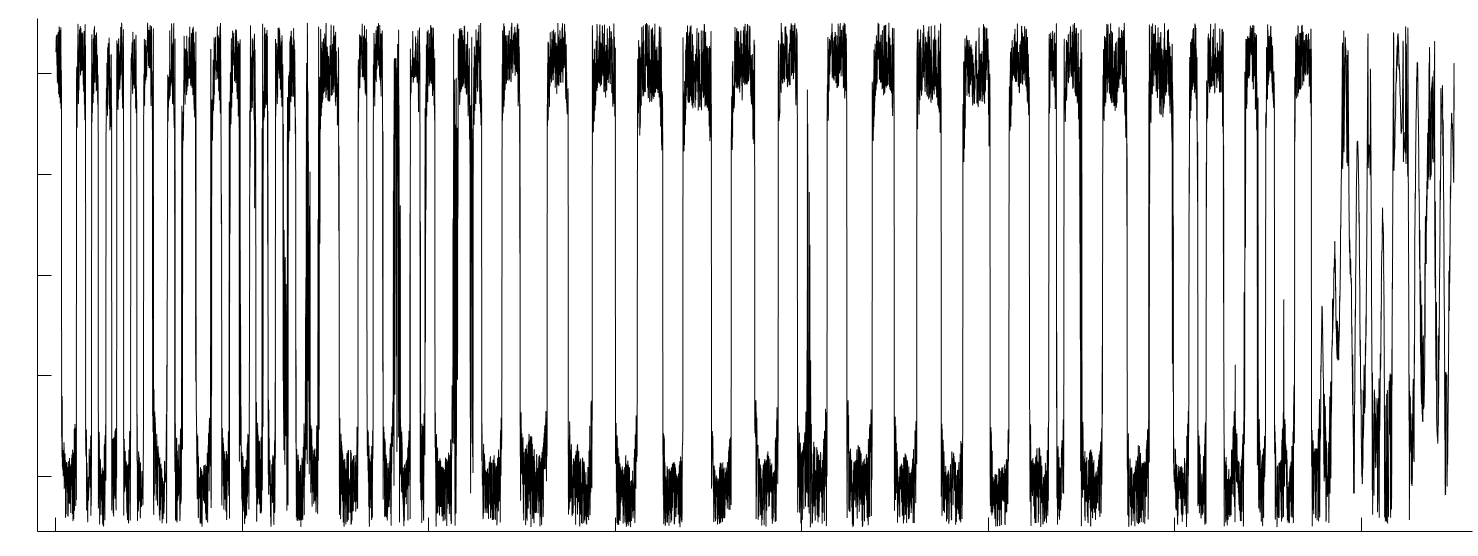}}
\put(58,120){\textbf{Signal generated via optimization alone}}
\put(125,-22){Sample number}
\put(-28,35){\rotatebox{90}{Signal value}}
\put(9.5,-5){0}
\put(80.5,-5){2000}
\put(159.5,-5){4000}
\put(238.5,-5){6000}
\put(-11,13.5){-0.4}
\put(-11,35){-0.2}
\put(0,56.5){0}
\put(-7,78){0.2}
\put(-7,99.5){0.4}
%\put(103,-5){100}
%\put(152,-5){150}
%\put(201,-5){200}
%\put(250,-5){250}
%\put(299,-5){300}
\end{picture}

\vspace{-1mm}
\begin{picture}(320,170)
\put(0,0){\includegraphics[width=110mm]{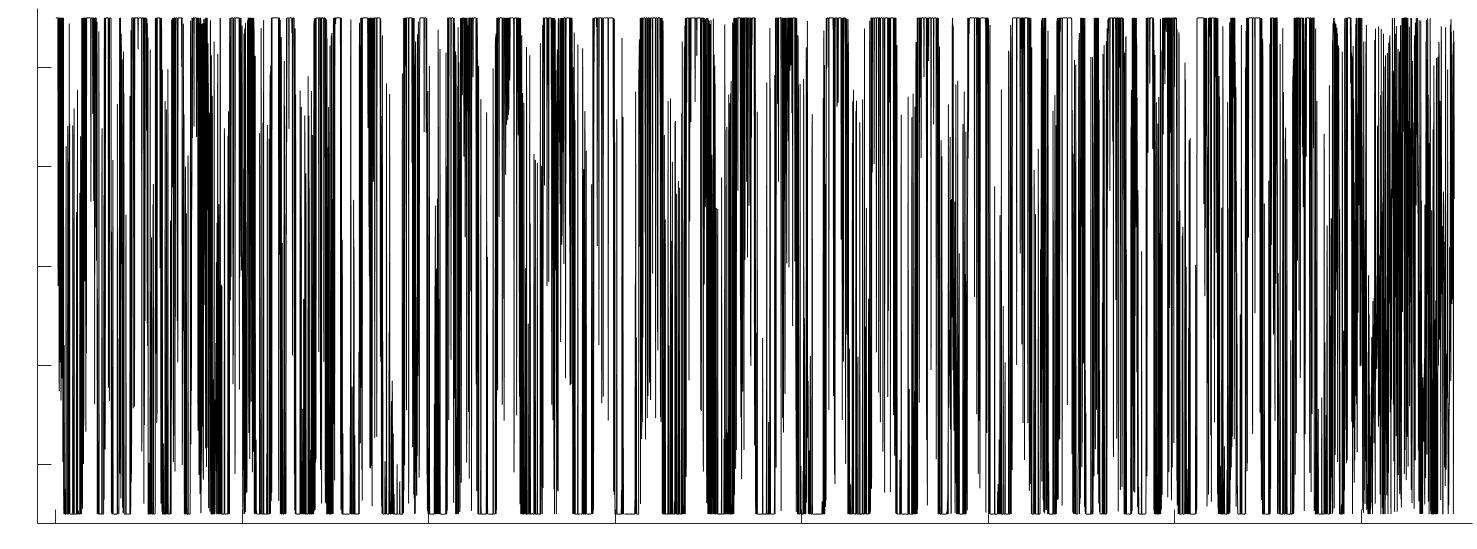}}
\put(4,120){\textbf{Signal generated via optimization plus stochastic interchange}}
\put(125,-22){Sample number}
\put(-28,35){\rotatebox{90}{Signal value}}
\put(9.5,-5){0}
\put(80.5,-5){2000}
\put(159.5,-5){4000}
\put(238.5,-5){6000}
\put(-11,13.5){-0.4}
\put(-11,35){-0.2}
\put(0,56.5){0}
\put(-7,78){0.2}
\put(-7,99.5){0.4}
\end{picture}

\vspace{8.5mm}

\caption{Comparison of signals generated via optimization on its own vs. optimization plus stochastic interchange. Both have the same autocorrelation (shown in Fig.~\ref{fig:autocorr}) to very high precision. However the signal generated by optimization on its own is clearly nonstationary; the temporal mixing by the stochastic interchange steps results in stationary generated signals.}
\label{fig:stationary}
\end{figure*}

\section{Results}
\label{sec:results}

\begin{figure*}[!t]

\centering

%\vspace{-14mm}
\vspace{-2mm}
\begin{picture}(320,220)
\put(0,0){\includegraphics[width=110mm]{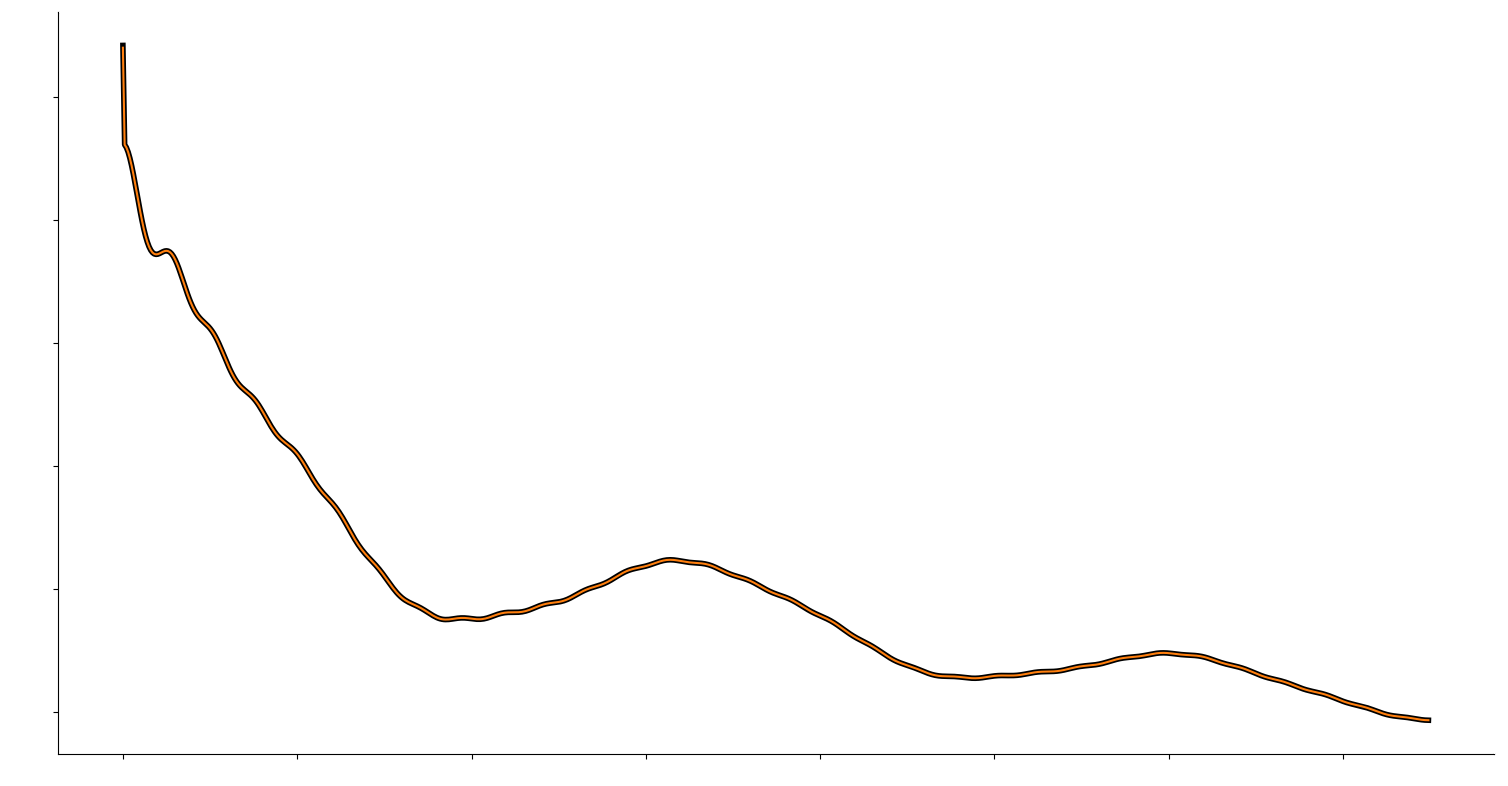}}
\put(50,170){\textbf{Autocorrelation of the generated signal}}
\put(98,-22){Delay (in number of samples)}
\put(23,-5){0}
\put(54,-5){100}
\put(90.5,-5){200}
\put(126.5,-5){300}
\put(163,-5){400}
\put(198.5,-5){500}
\put(235,-5){600}
\put(271.5,-5){700}
\put(-13,12){-0.05}
\put(3,37.5){0}
\put(-10,63){0.05}
\put(-10,88.5){0.10}
\put(-10,114){0.15}
\put(-10,139.5){0.20}
\put(-28,58){\rotatebox{90}{Correlation}}
\end{picture}

\vspace{9mm}

\caption{Autocorrelation function of a signal generated by optimization plus stochastic interchange. The black line is the desired autocorrelation, the orange line on top is the autocorrelation of the generated signal. There is near-perfect agreement; the autocorrelation of the generated signal accounts for 99.998\% of the variance of the desired autocorrelation.}
\label{fig:autocorr}
\end{figure*}
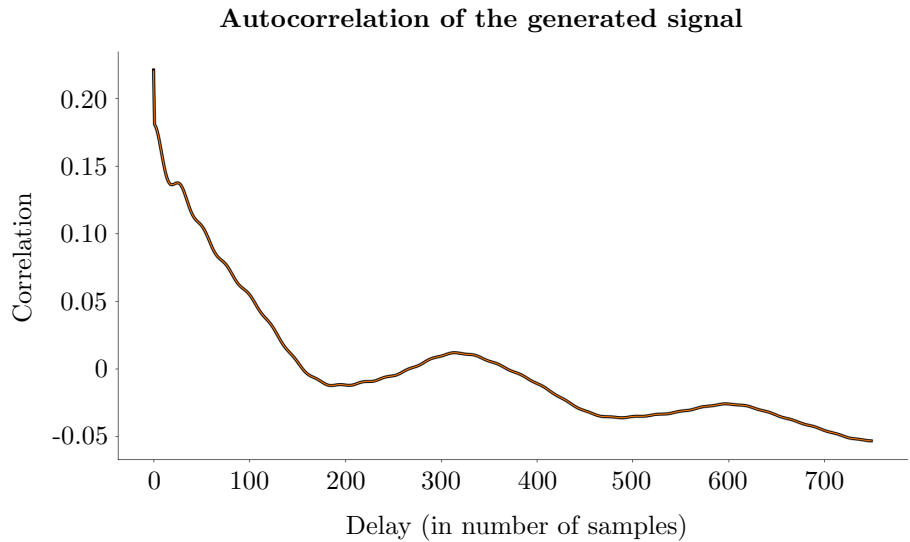

\begin{figure*}[!t]
\centering

\vspace{15mm}
\begin{picture}(320,200)
\put(0,0){\includegraphics[width=110mm]{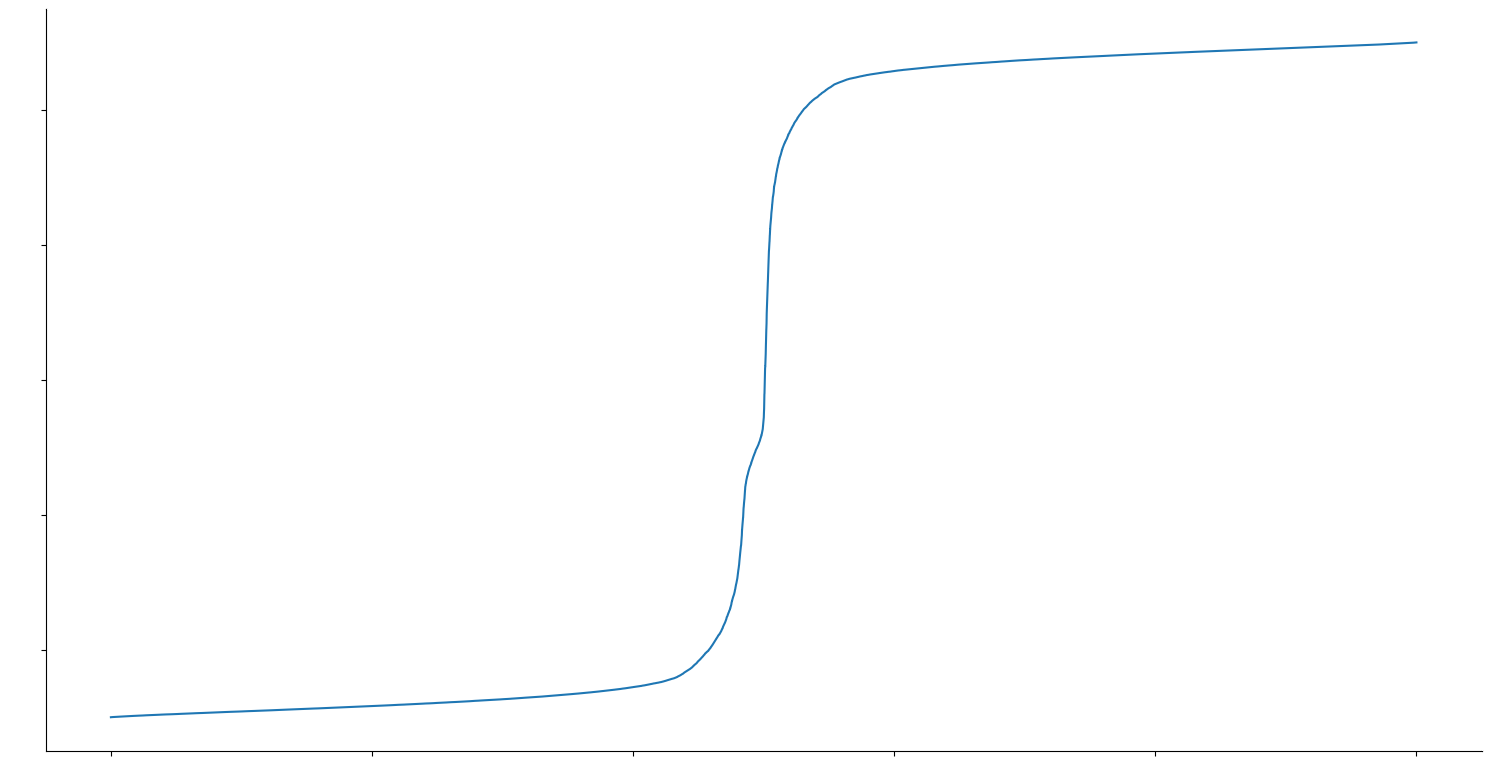}}
\put(12,170){\textbf{Cumulative distribution function of the generated signal}}
\put(125,-22){Sample number}
\put(-28,58){\rotatebox{90}{Sample value}}
\put(21,-5){0}
\put(66,-5){10000}
\put(120.5,-5){20000}
\put(175,-5){30000}
\put(230,-5){40000}
\put(284.5,-5){50000}
\put(-11,24){-0.4}
\put(-11,52){-0.2}
\put(0,80){0}
\put(-7,109){0.2}
\put(-7,137){0.4}
\end{picture}

\vspace{9mm}

\caption{The CDF of the generated signal shows that it is nearly a binary stochastic signal, that is, most samples take on an extreme value in the allowed range. This arises naturally from the optimization procedure when we constrain the values of the signal to lie within a certain range while specifying a large signal power when choosing the power spectral density, or equivalently the autocorrelation function.}
\label{fig:cdf}
\end{figure*}
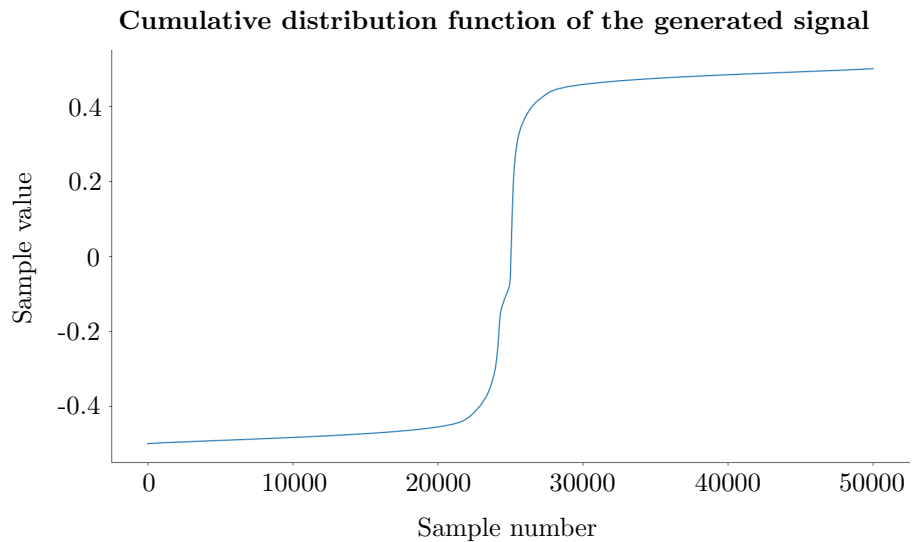

We find that interspersing optimization steps with stochastic interchange resolves issues with either method on its own:
\begin{enumerate}
\item It converges faster than stochastic interchange; generating a signal of length $n$ can take as little as $\mathcal{O}(n \: \text{log} \: n)$ time without parallelization or $\mathcal{O}(n)$ time with full parallelization. See Fig.~\ref{fig:convergence} for a comparison of convergence rates.
\item It has greater flexibility than stochastic interchange due to the use of optimization, allowing the user to specify constraints on the PDF rather than specifying an exact PDF. An example of the resulting distribution of signal values can be seen in Fig.~\ref{fig:cdf}. This additional flexibility enables convergence to an extremely tight agreement between desired and attained autocorrelation as can be seen in Figs.~\ref{fig:convergence}~and~\ref{fig:autocorr}.
\item It escapes quickly from local optima due to the stochastic interchange steps.
\item It achieves good stationarity due to the stochastic interchange steps, as can be seen in Fig.~\ref{fig:stationary}.
\end{enumerate}

%%%%% %%%%% %%%%% %%%%% %%%%% %%%%% %%%%% %%%%% %%%%% %%%%% %%%%% %%%%%

\section{Discussion}
%%%%% %%%%% %%%%% %%%%% %%%%% %%%%% %%%%% %%%%% %%%%% %%%%% %%%%% %%%%%

This paper has introduced a novel algorithm for generating stationary stochastic signals with a specified autocorrelation and a constrained PDF. This algorithm is an improvement on Hunter and Kearney's stochastic interchange method~\citep{hunter1983generation}, obtained by interspersing gradient optimization steps. This improves its best-case runtime from $\mathcal{O}(n^2)$ to $\mathcal{O}(n \: \text{log} \: n)$ without parallelization or $\mathcal{O}(n)$ with full parallelization, while also improving its accuracy and flexibility. 

\vspace{1.2mm}
While the problem of generating stationary stochastic signals with a specified autocorrelation and PDF has largely been solved by methods involving nonlinear filtering, the stochastic interchange method is uniquely positioned to generalize to new settings such as generating 2-dimensional stochastic signals with a specified 2D autocorrelation and PDF, or generating stochastic signals with specified higher-order autocorrelation functions and PDF.

\vspace{1.2mm}
The authors have already found this work to be useful in generating experimental stimuli in the context of nonlinear system identification~\citep{spanbauer2020coarse}. We hope that this work and its generalizations to stochastic signals with more broadly varying statistics will prove useful for generating experimental stimuli for optimal system identification, as well as for the simulation of non-Gaussian physical processes.

\newpage

%%%%% %%%%% %%%%% %%%%% %%%%% %%%%% %%%%% %%%%% %%%%% %%%%% %%%%% %%%%%
\section*{Acknowledgements}
%%%%% %%%%% %%%%% %%%%% %%%%% %%%%% %%%%% %%%%% %%%%% %%%%% %%%%% %%%%%

The authors would like to thank Fonterra for supporting this research.

%%%%% %%%%% %%%%% %%%%% %%%%% %%%%% %%%%% %%%%% %%%%% %%%%% %%%%% %%%%%

\bibliography{mybib}

\end{document}